\documentstyle[prl,aps,epsf]{revtex}

\begin{document}
\draft
\title{Comment on ``Singularities in axially symmetric solutions of
Einstein-Yang Mills and related theories, by Ludger Hannibal, 
[hep-th/9903063]'' }
\author{
{\bf Burkhard Kleihaus}}
\address{NUIM, Department of Mathematical Physics, Maynooth, Co. Kildare,
Ireland\\
}
\author{and\\}
\author{
{\bf Jutta Kunz}}
\address{Fachbereich Physik, Universit\"at Oldenburg, Postfach 2503,
D-26111 Oldenburg, Germany
\\
}
\date{\today}
\maketitle
\bigskip
\begin{abstract}
We point out that the statements in [hep-th/9903063] concerning the
regularity of static axially symmetric solutions in Yang-Mills-dilaton (YMD)
\cite{YMD}
and Einstein-Yang-Mills(-dilaton) (EYMD) theory \cite{EYMD,an} are incorrect,
and
that the non-singular local gauge potential of the YMD solutions \cite{regu}
is twice differentiable.

\end{abstract}
\bigskip

We have constructed numerically static axially symmetric solutions
in Yang-Mills-dilaton (YMD) \cite{YMD} and
Einstein-Yang-Mills-dilaton (EYMD) theory \cite{EYMD,an},
employing a singular form of the gauge potential. 
For the solutions of YMD theory we have recently demonstrated explicitly,
that the singular gauge potential can be locally gauge transformed into 
a well defined gauge potential \cite{regu}.

After that, Hannibal has claimed in his paper \cite{LH},
``We show that the solutions of $SU(2)$ Yang-Mills-dilaton and
Einstein-Yang-Mills-dilaton theories described in a sequence
of papers by Kleihaus and Kunz are not regular in the gauge field part''(*).
Here we comment on this paper \cite{LH} only as far as our work 
\cite{YMD,EYMD,an,regu} is concerned.
\bigskip

In \cite{LH} static axially symmetric Ans\"atze for the 
$SU(2)$ gauge potential \cite{RR,YMD,EYMD,an}
have been considered and regularity conditions 
for the gauge field functions parameterizing the 
Ansatz \cite{an} have been derived \cite{regcon}. 
Comparing the properties of the gauge field functions
employed in \cite{YMD,EYMD,an} to these regularity conditions, 
it was then concluded in \cite{LH}, 
that ``the solutions constructed by Kleihaus and Kunz \cite{YMD,EYMD,an} 
do not have the regular form''.

While the solutions constructed numerically in 
\cite{YMD,EYMD,an} are presented within a singular form of the gauge potential
in the sense,
that the gauge potential is not well defined on the $z$-axis and
at the origin \cite{regu},
this, however, only means that regularity of the solutions 
is not guaranteed a priori.
It does not mean, that the solutions are not regular.

Since the gauge potential transforms under gauge transformations,
a regular gauge potential $\hat{A}_\mu$ can be 
gauge transformed into a singular gauge potential $A_\mu$  
by a singular gauge transformation. 
Both gauge potentials would describe the same physical solution.
From the observation that the gauge potential $A_\mu$ is in a 
singular form, one could neither conclude
that there is no regular gauge potential $\hat{A}_\mu$ 
nor, in particular, that the physical solution is not regular.
Therefore, the claim (*) in the abstract of Hannibals paper 
\cite{LH} is not correct and misleading.

\bigskip

Hannibal \cite{LH} has then considered the gauge transformation \cite{regu},
which transforms locally the singular gauge potential of YMD solutions 
\cite{YMD} into a non-singular gauge potential.
He recognizes, that the transformed gauge potential is continuous at 
$\theta =0$ (i.~e.~at $x=y=0$), but he claims that
``the potentials are still possibly not differentiable at $\theta =0$''.

However, it is a trivial task to check that the transformed
gauge potential \cite{regu} is differentiable.
For instance for winding number $n=2$ and $z>0$,  
the non-singular gauge potential
is given explicitly in \cite{regu} in Cartesian coordinates,

\begin{eqnarray}
\hat{A}_x &=&  
-\frac{x}{12r^4}\left[r\partial_r\tilde{H}_{12} 
- 3\tilde{H}_{12}\right]  \rho^2 \tau^2_\varphi 
-\frac{y}{6r^4}\tilde{H}_{43} \rho^2 \tau^2_\rho 
+\frac{y}{2r^2}\left[f^2+2g-1\right] \tau_3 \ ,
\nonumber\\
\hat{A}_y &=&  
-\frac{y}{12r^4}\left[r\partial_r\tilde{H}_{12} 
- 3\tilde{H}_{12}\right]  \rho^2 \tau^2_\varphi 
+\frac{x}{6r^4}\tilde{H}_{43} \rho^2 \tau^2_\rho 
-\frac{x}{2r^2}\left[f^2+2g-1\right] \tau_3 \ ,
\nonumber\\
\hat{A}_z &=&  
\frac{1}{4r^3} \tilde{H}_{12}\rho^2 \tau^2_\varphi  \ ,
\label{A_lo}
\end{eqnarray}
expanded to lowest order in the variable $\rho=\sqrt{x^2+y^2}$.
Using $\rho^2 \tau^2_\varphi = -2xy \tau_1 +(x^2-y^2) \tau_2$ and 
$\rho^2 \tau^2_\rho = (x^2-y^2)  \tau_1 + 2xy \tau_2$, one sees immediately,
that the gauge potential (\ref{A_lo}) is differentiable at 
$x=y=0$. 
\footnote{
Note, that the functions $\tilde{H}_{12},\tilde{H}_{43},f,g$ \cite{regu}
and their derivatives are bounded functions of $r=\sqrt{x^2+y^2+z^2}$.
The function $f$ used here and in \cite{regu}
should not be confused with the function $f$ in \cite{LH}.}

In addition it is evident, that the gauge potential is twice
differentiable at $x=y=0$, provided one can show that no functions 
like $\rho x$, $\rho y$ arise from the next to leading order terms.
But this is straightforward, since it concerns only the function
$\hat{F}_4$, multiplying the matrix $\tau_3$, see \cite{regu}.
Observing from \cite{regu},
that the functions $H_3$ and $\Gamma_{(z)}^{(n)}$ are
odd in $\theta$ up to order $\theta^3$ (inclusively) while 
the function $H_4$ is even in $\theta$ up to order $\theta^2$, 
it is easy to see,
that the function $\hat{F}_4$ is odd in $\theta$ up to order $\theta^3$.
Thus the next to leading order term vanishes, indeed. 
At the origin a similar consideration shows that
the next to leading order terms only contribute 
functions which are twice differentiable at the origin.
Consequently, the gauge transformed gauge potential is twice differentiable,
even in the lowest order of the expansion \cite{regu}. 

Nevertheless, we have carried out the expansion near the positive $z$-axis 
to the next order and have found 
that the next to leading order terms vanish for all gauge field functions 
for the YMD solutions \cite{YMD} with winding number $n=2,3$
\cite{regu1}.
\bigskip

To conclude, it has not been shown in \cite{LH}
that the static axially symmetric solutions 
constructed numerically in \cite{YMD,EYMD,an} are not regular.
Furthermore,
the non-singular gauge potentials obtained to lowest order in \cite{regu}
are continuous, differentiable and twice differentiable 
at $x=y=0$ and at the origin. 
Consequently, the singular gauge potential 
of the solutions obtained in \cite{YMD}
can locally be gauge transformed into regular form.


\begin{thebibliography}{000}
\bibitem{regcon}
This has been done before \cite{RR}
for the original parameterization of the Ansatz, leading 
to the same result.
\end{thebibliography}
\end{document}